\documentclass[prd,showpacs,twocolumn,preprintnumbers]{revtex4}

\usepackage{amsfonts,amsmath,graphicx,bm}

\newcommand{\ms}{\;\;\:}

\begin{document}

\title{The bottomonium spectrum from lattice QCD with 2+1 flavors of\\ domain wall fermions}

\author{Stefan Meinel}
\email[E-mail: ]{S.Meinel@damtp.cam.ac.uk}
\affiliation{Department of Applied Mathematics and Theoretical Physics,
University of Cambridge,
Centre for Mathematical Sciences,
Cambridge CB3 0WA, United Kingdom}
\date{April 21, 2009}

\pacs{12.38.Gc, 14.40.Gx}
\preprint{DAMTP-2009-25}

\begin{abstract}
Recently, realistic lattice QCD calculations with 2+1 flavors of domain wall fermions and the Iwasaki gauge action
have been performed by the RBC and UKQCD collaborations. Here, results for the bottomonium spectrum computed
on their gauge configurations of size $24^3\times64$ with a lattice spacing of approximately $0.11$~fm and four different values
for the light quark mass are presented. Improved lattice NRQCD is used to treat the $b$ quarks inside the bottomonium. The results for the radial
and orbital energy splittings are found to be in good agreement with experimental measurements, indicating that systematic errors are small.
The calculation of the $\Upsilon(2S)-\Upsilon(1S)$ energy splitting provides an independent determination of the
lattice spacing. For the most physical ensemble it is found to be $a^{-1}=1.740(25)(19)$ GeV,
where the first error is statistical/fitting and the second error is an estimate of the systematic errors due to the
lattice NRQCD action.
\end{abstract}

\maketitle

\section{Introduction}

Bottomonium mesons, the bound states of bottom quark-antiquark pairs, play an important role in the study of the
strong interactions. The spectrum of bottomonium is known very well from experiment and there are many different approaches
to calculating it theoretically. Lattice QCD provides a model-independent and accurate way of doing this.

One of the most important steps toward realistic lattice QCD calculations was the inclusion of dynamical
light ($u$, $d$ and $s$) sea quarks. For many quantities of phenomenological interest, lattice QCD now
allows reliable non-perturbative calculations that were previously impossible. In order to further
control systematic errors and increase the confidence in lattice results, it is crucial to consider several
different lattice actions and thereby test universality.

The RBC and UKQCD collaborations have recently started large-scale lattice QCD calculations
\cite{Antonio:2006px,Allton:2007hx,Antonio:2008zz,Boyle:2007qe,Antonio:2007pb,Allton:2008pn} with
dynamical domain wall fermions and renormalization-group-improved gauge actions.
The domain wall fermion action \cite{Kaplan:1992bt,Shamir:1993zy,Furman:1994ky} has an approximate chiral
symmetry that becomes exact, even at finite lattice spacing, when the extent $L_s$ of the auxiliary
fifth dimension is taken to infinity. This leads to better control over the renormalization of operators,
a reduction of discretization errors and more reliable chiral extrapolations.

The gauge configurations created by the RBC and UKQCD collaborations have been made publicly available.
In this work, the ensembles of size $24^3\times 64$, $L_s=16$ described in detail in \cite{Allton:2008pn}
were used to compute the spectrum of bottomonium. Chiral symmetry is not as important for bottomonium
as it is for light hadrons, but the calculations presented here nevertheless provide useful tests of
the recent lattice calculations by the RBC and UKQCD collaborations. In particular, they provide
an independent determination of the lattice spacing and a good way of tuning the $b$ quark mass.
The value of the bare $b$ quark mass obtained here can also be used in lattice QCD calculations for heavy-light
hadrons such as $B$ mesons, which are of considerable importance for the phenomenology of weak decays
and tests of the Standard Model.

The $b$ quark has a mass larger than the inverse lattice spacing, so that the standard lattice actions as used for
light quarks are not suitable to describe it. A preliminary calculation of the bottomonium spectrum on the RBC/UKQCD
gauge configurations using a relativistic heavy-quark action was presented in \cite{Li:2008kb}. Here, improved
non-relativistic QCD (NRQCD) \cite{Thacker:1990bm, Lepage:1992tx, Davies:1994mp} is employed instead,
following closely the methods used in the calculation of the bottomonium spectrum on the MILC gauge
configurations in \cite{Gray:2005ur} (these configurations use an improved staggered fermion action for the sea quarks
and a one-loop Symanzik-improved gluon action).

The lattice calculation and analysis methods are described in detail in Sec.~\ref{sec:lattice_calculation}. The results for the
tuning of the $b$ quark mass and tests of the dispersion relation are given in Sec.~\ref{sec:mass_tuning} and
\ref{sec:speed_of_light}, respectively. In Sec.~\ref{sec:radial_orbital_splittings}, the results for the radial and orbital
energy splittings as well as the lattice spacing are presented, followed by the fine and hyperfine structure in
Sec.~\ref{sec:spin_dep_splittings}.

\section{\label{sec:lattice_calculation}The lattice calculation}

\subsection{Lattice actions and parameters}

The details of the domain wall fermion and Iwasaki gauge actions as used by the RBC and UKQCD collaborations
are given in \cite{Antonio:2006px}. Here, the gauge configurations of size $24^3\times 64$
as described in \cite{Allton:2008pn} were used. These have $L_s=16$, $\beta=2.13$ and the strange quark mass is $am_s=0.04$.
There are ensembles with four different values for the degenerate light (up and down) quark mass $a m_l$, as shown in Table
\ref{tab:lattices}.
\begin{table}
\begin{tabular}{lcccccr}
\hline\hline
$am_l$ & \hspace{2ex} & $u_{0L}$ & \hspace{2ex} & MD range (step)   & \hspace{2ex} & $n_{\rm conf}$ \\
\hline
$0.005$   && $0.8439$ && $915$ - $8665$ $(25)$ && $311$          \\
$0.01$    && $0.8439$ && $1475$ - $8525$ $(25)$ && $283$          \\
$0.02$    && $0.8433$ && $1800$ - $3600$ $(25)$ && $73$           \\
$0.03$    && $0.8428$ && $1275$ - $3050$ $(25)$ && $72$           \\
\hline\hline
\end{tabular}
\caption{\label{tab:lattices}The ensembles of RBC/UKQCD gauge configurations used here.}
\end{table}
The ``measurements'' in this work were started at the same molecular dynamics (MD) time as in \cite{Allton:2008pn}
to ensure complete thermalization. Note however that the $a m_l=0.005$ and $a m_l=0.01$ ensembles have since been
extended and the additional configurations were included here. Measurements were performed
every 25 steps of MD time, as discussed further in Sec.~\ref{sec:autocorr}.

The lattice NRQCD action for the $b$ quark is the same as in the previous study of the bottomonium spectrum in \cite{Gray:2005ur},
with stability parameter $n=2$. The full details of the action can be found in e.g.~\cite{Dalgic:2003uf}. After the initial tuning,
which will be described in Sec.~\ref{sec:results}, the bare $b$ quark mass was set to $a m_b=2.536$.

All couplings in the NRQCD action are set to their tree-level values, but the action is tadpole-improved
\cite{Lepage:1992xa}, which accounts for a large amount of the renormalization.
The mean link in Landau gauge $u_{0L}$ is used as the tadpole improvement parameter;
the values of $u_{0L}$ for the different ensembles are listed in Table \ref{tab:lattices}.

The NRQCD action in use includes relativistic correction terms up to order $\mathcal{O}(v^4)$
where $v$ is the internal speed of the $b$ quarks inside the meson. For bottomonium, one has $v^4\approx0.01$.
The action is also tree-level Symanzik improved. Systematic errors depend strongly on the observable under consideration,
and will be discussed individually in Sec.~\ref{sec:results}. However, finite-volume errors are expected to be
negligible in all cases due to the small size of the bottomonium mesons.

\subsection{\label{sec:2ptcalc}Calculation of the meson two-point functions}

As in \cite{Davies:1994mp}, the heavy-heavy meson correlators with momentum $\bm{p}$ were computed from
\begin{eqnarray}
\nonumber&\hspace{-2.5ex}&C(\Gamma_{\rm sk}, \Gamma_{\rm sc}, \bm{p}, t-t')\\
\nonumber&\hspace{-2.5ex}&=\!\sum_{\bm{x}_1,\bm{x}_2}\!\mathrm{Tr}
\left[ G^\dag
(\bm{x}_1,t,\:\bm{x}'_1,t')\Gamma_{\rm sk}^\dag(\bm{x}_1\!-\!\bm{x_2})
\tilde{G}_{\rm sc} (\bm{x}_2,t,\:\bm{x}'_1,t') \right]\\
&\hspace{-2.5ex}& \hspace{6ex} \times \: e^{-i\bm{p}\frac{\bm{x}_1+\bm{x}_2}{2}}
\label{eqn:lat_corr}
\end{eqnarray}
where $t>t'$ and
\begin{eqnarray}
\nonumber\tilde{G}_{\rm sc}(\bm{x}_2,t,\:\bm{x}'_1,t')
&=&\sum_{\bm{x}'_2}
G(\bm{x}_2,t,\:
\bm{x}'_2,t')\Gamma_{\rm sc}(\bm{x}'_1\!-\bm{x}'_2)\\
&&\hspace{4ex}\times\:e^{i\bm{p}\frac{\bm{x}'_1+\bm{x}'_2}{2}}.
\label{eqn:initial_green}
\end{eqnarray}
In Eqs.~(\ref{eqn:lat_corr}) and (\ref{eqn:initial_green}), which are understood to be for a single gauge configuration,
$G$ denotes the heavy-quark propagator which is $2\times2$ matrix-valued in spinor space and $3\times3$ matrix-valued in color space. The
functions $\Gamma_{\rm sc/sk}$ are the ``smearing functions'' at source and sink, respectively, which are also $2\times2$ matrix-valued
in spinor space. No gauge links were included in $\Gamma_{\rm sc/sk}$; instead the gauge configurations were fixed to Coulomb
gauge.

In Table \ref{tab:smear_funcs} the bottomonium states considered in this work are listed, together with their continuum
quantum numbers, smearing functions $\Gamma(\bm{r})$ and representations of the octahedral group \cite{Johnson:1982yq}.
\begin{table*}
\begin{ruledtabular}
\begin{tabular}{llllllllllllcll}
 \\[-3ex]
Name & \hspace{2ex} & $L$ & \hspace{2ex} & $S$ & \hspace{2ex} & $J$ & \hspace{2ex} & $P$ & \hspace{2ex} & $C$ & \hspace{2ex} & Lattice rep. $\mathcal{R}^{PC}$  & \hspace{2ex} & $\Gamma(\bm{r})$ \\
 \\[-3ex]
\hline
 \\[-3ex]
$\eta_b(nS)$ && 0 && 0 && 0 && $-$ && $+$ && $A_1^{-+}$ && $\phi_{nS}(\bm{r})$ \\
 \\[-3ex]
$\Upsilon(nS)$ && 0 && 1 && 1 && $-$ && $-$ && $T_{1}^{--}$ && $\phi_{nS}(\bm{r})\:\sigma^i$ \\
 \\[-3ex]
$h_b(nP)$ && 1 && 0 && 1 && $+$ && $-$ && $T_{1}^{+-}$ && $\phi_{nP}(\bm{r})\:r^i/r_0$ \\
 \\[-3ex]
$\chi_{b0}(nP)$ && 1 && 1 && 0 && $+$ && $+$ && $A_1^{++}$ && $\phi_{nP}(\bm{r})\:(\mathbf{r}\cdot\boldsymbol{\sigma})/r_0$ \\
 \\[-3ex]
$\chi_{b1}(nP)$ && 1 && 1 && 1 && $+$ && $+$ && $T_1^{++}$ && $\phi_{nP}(\bm{r})\:(\bm{r}\times\bm{\sigma})^i/r_0$ \\
 \\[-3ex]
$\chi_{b2}(nP)$ && 1 && 1 && 2 && $+$ && $+$ && $T_2^{++}$ && $\phi_{nP}(\bm{r})\:(r^i\sigma^j+r^j\sigma^i)/r_0$ \hfill {\scriptsize$(i\neq j)$}  \\
 \\[-3ex]
$\eta_{b}(nD)$ && 2 && 0 && 2 && $-$ && $+$ && $T_2^{-+}$ && $\phi_{nD}(\bm{r})\:r^i r^j/r_0^2$ \hfill {\scriptsize$(i\neq j)$} \\
 \\[-3ex]
$\Upsilon_{2}(nD)$ && 2 && 1 && 2 && $-$ && $-$ && $E^{--}$ && $\phi_{nD}(\bm{r})\:(r^i r^j \sigma^k-r^j r^k\sigma^i)/r_0^2$ \hspace{2ex} {\scriptsize$(i\neq j,\:\: k\neq j)$} \\
 \\[-3ex]
\end{tabular}
\end{ruledtabular}
\caption{\label{tab:smear_funcs}The smearing functions $\Gamma(\bm{r})$. See e.g.~\cite{Johnson:1982yq} for the
irreducible representations of the octahedral group.}
\end{table*}
As can be seen in the table, all representations are chosen to be different, so that no mixing is expected here.
The radial functions $\phi_{nS}(\bm{r})$, $\phi_{nP}(\bm{r})$ and $\phi_{nD}(\bm{r})$ for the $n$-th radially
excited $S$-wave ($L=0$), $P$-wave ($L=1$) and $D$-wave ($L=2$) states were taken from the
corresponding hydrogen atom wave functions and are given in Table \ref{tab:radial_funcs}.
\begin{table}
\begin{tabular}{lccl}
\hline\hline
 \\[-3ex]
State &&& $\phi(\bm{r})$ \\
 \\[-3ex]
\hline
 \\[-3ex]
$1S$ &&& $\exp[-|\bm{r}|/r_0]$ \\
 \\[-3ex]
$2S$ &&& $\left[1-|\bm{r}|/(2r_0)\right]\:\exp[-|\bm{r}|/(2r_0)]$ \\
 \\[-3ex]
$3S$ &&& $\left[1-2|\bm{r}|/(3r_0)+2|\bm{r}|^2/(27r_0^2)\right]\:
\exp[-|\mathbf{r}|/(3r_0)]$ \\
 \\[-3ex]
$1P$ &&& $\exp[-|\mathbf{r}|/(2r_0)]$ \\
 \\[-3ex]
$2P$ &&& $\left[1-|\mathbf{r}|/(6r_0)\right]\exp[-|\mathbf{r}|/(3r_0)]$ \\
 \\[-3ex]
$1D$ &&& $\exp[-|\mathbf{r}|/(3r_0)]$ \\
 \\[-3ex]
\hline\hline
\end{tabular}
\caption{\label{tab:radial_funcs}The radial functions $\phi(\bm{r})$}
\end{table}
The same lattice representations were used at source and sink but the radial smearing functions
were allowed to be different. The smearing parameters $r_0$ (in lattice units) were set to $1.0$ ($1S$), $0.8$ ($2S$),
$0.6$ ($3S$), $0.5$ ($1P$), $0.4$ ($2P$) and $0.5$ ($1D$), respectively.

Note that $\tilde{G}$ in Eq.~(\ref{eqn:initial_green}) can be computed efficiently
by using the function
\begin{equation}
\Gamma_{\rm sc}(\bm{x}'_1\!-\bm{x}'_2)\:e^{i\bm{p}\frac{\bm{x}'_1+\bm{x}'_2}{2}}
\end{equation}
as the initial condition in the heavy quark evolution equation. For
some states it is computationally more convenient to remove the Pauli matrix in
$\Gamma_{\rm sc}$ in the initial condition, and instead including it explicitly in the trace
in Eq.~(\ref{eqn:lat_corr}). In this way, different spin directions can be obtained with
a single $\tilde{G}$.

Finally, note that on a finite lattice, the smearing functions must satisfy the periodic
boundary conditions. This was ensured by setting the smearing functions to zero outside a ball
with radius $R$ smaller than half the spatial lattice dimension. In this way, the wrapping around the
lattice boundaries does not cause any problems. Since the smearing functions decay exponentially with
the separation between quark and antiquark, $R$ can be chosen such that the important features remain.
To ensure symmetry, the same cut-off radius must be taken at source and sink.

In order to increase statistics, the correlators (\ref{eqn:lat_corr}) were averaged over eight different spatial origins
$\bm{x}_1'$ located at the corners of a cube with side length $L/2=12$. In addition, four different source time
slices $t'$ with an equal spacing of $16$ were used, thus leading to a total of 32 origins per configuration.
Furthermore, the locations of the origins on the lattice were shifted randomly from configuration to configuration
in order to decrease autocorrelations.

\subsection{\label{sec:fitting_and_analysis}Fitting and analysis details}

\subsubsection{Bayesian multi-exponential fitting}

After choosing a set of smearing functions $\Gamma(\bm{r})$ with equal lattice representations
but different radial functions (e.g.~$1S$, $2S$ and $3S$), the square
matrix of correlators obtained by taking all combinations for source and sink was computed.

The matrix of correlators $\langle C(\Gamma_{\rm sk}, \Gamma_{\rm sc}, \bm{p}, t-t') \rangle$
was simultaneously fitted by a function of the form
\begin{equation}
\sum_{n=0}^{n_{\rm exp}-1} A_n(\Gamma_{\rm sc})\:A^*_n(\Gamma_{\rm sk}) \:e^{-E_n(t-t')}
\label{eqn:fitfunction}
\end{equation}
where $E_n$ is the energy of $n$-th state
and $A_n(\Gamma)$ is the (real) amplitude for this state
to be created by the operator with smearing function $\Gamma(\bm{r})$.

To ensure the correct ordering of the states in terms of their energy,
the fit parameters were actually chosen to be the logarithms of the energy
differences between neighboring states (in lattice units)
\begin{equation}
 \ln(E_{n+1}-E_n)
\end{equation}
and the logarithm of the ground state energy, $\ln(E_0)$. Furthermore, the
amplitudes for the excited states were written as
\begin{equation}
 A_n(\Gamma)=A'_n(\Gamma) A_0(\Gamma),
\end{equation}
taking the relative amplitudes $A'_n(\Gamma)$ (for $n\geq1$) and the ground state
amplitude $A_0(\Gamma)$ as the fit parameters.

The Bayesian fitting method described in \cite{Lepage:2001ym} was used, where the $\chi^2$
function is augmented by
\begin{equation}
 \chi^2 \: \rightarrow \: \chi^2+\chi^2_{\rm prior}
\end{equation}
with the Gaussian prior
\begin{equation}
\chi^2_{\rm prior}=\sum_i\frac{(p_i-\tilde{p}_i)^2}{\sigma_{\tilde{p}_i}^2}. \label{eq:priors}
\end{equation}
Here, $\{ p_i \}=\{ A_0(\Gamma),\: A'_n(\Gamma),\: \ln(E_0),\: \ln(E_{n+1}-E_n) \}$
are the fitting parameters, and the prior for each parameter $p_i$ is given by its central value
$\tilde{p}_i$ and width $\sigma_{\tilde{p}_i}$.

\begin{figure}
 \includegraphics[width=0.9\linewidth]{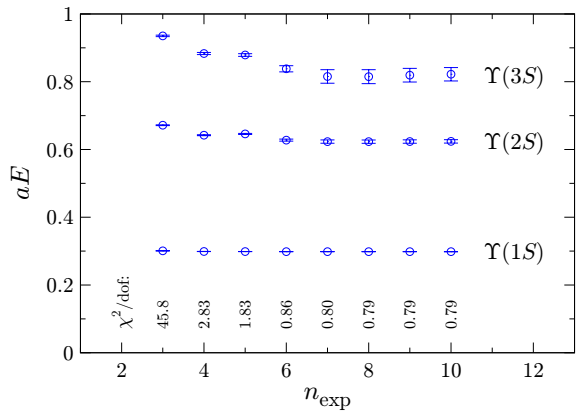}
 \caption{\label{fig:n_exp_convergence} Fit results for the $3\times 3$
matrix correlator with the $\left\{ \Upsilon(1S),\: \Upsilon(2S),\: \Upsilon(3S)\right\}$ smearings
as a function of the number of exponentials. The values of $\chi^2$ divided by the number of degrees of freedom are also shown.
The results are for the ensemble with $am_l=0.005$.}
\end{figure}

The Bayesian method allows the inclusion of an arbitrary number of exponentials $n_{\rm exp}$ in
(\ref{eqn:fitfunction}) and hence the fitting in the full range of Euclidean time
$t-t'$ between source and sink. Here, only the points with $t-t'=0$ were excluded in the fits. The number of
exponentials is increased until the fit results and error estimates become
independent of $n_{\rm exp}$. This is demonstrated in Fig.~\ref{fig:n_exp_convergence}.

In the following discussion of the prior choices $\{ \tilde{p}_i, \sigma_{\tilde{p}_i} \}$
we will distinguish between parameters for \emph{low-lying}
and \emph{high-lying} states. For example, in a $3\times3$ matrix fit containing sources optimized for
the $\Upsilon(1S)$, $\Upsilon(2S)$ and $\Upsilon(3S)$ states, these three states will be referred to as
low-lying, as their energies and amplitudes will be well-determined by the data, while higher excitations
will be referred to as high-lying.

The prior widths for the parameters of the low-lying states were chosen to be about 10 times larger
than the resulting error estimates from the fit. This ensures that the influence of the priors on these
parameters is negligible. Initial guesses for the central values were obtained from unconstrained fits
including only a small number of exponentials at large Euclidean time.

For the high-lying states, the priors for the logarithms of the energy-splittings between successive states,
$\ln(E_{n+1}-E_n)$, were set to $-1.4$ (corresponding to about 400 MeV) with a width of 1.
The priors for the relative amplitudes $A'_n(\Gamma)$ of the high-lying states were set to zero with a
width of 5.

\subsubsection{NRQCD energy shift}

Due to the use of lattice NRQCD, where the heavy quark mass has been integrated out of the action, all
energies obtained from the fits are shifted,
\begin{equation}
 E = E_{\rm phys} - 2\:C
\end{equation}
where $C$ is approximately equal to the heavy quark mass. Since $C$ is the same for all states,
energy splittings are unaffected by this shift.

The physical mass of a meson (and hence $C$) can be calculated from the energy
difference between its states with $\bm{p}=0$ and $\bm{p}\neq0$. Assuming the relativistic
continuum dispersion relation
\begin{equation}
 E(\bm{p}) = \sqrt{ \bm{p}^2 + M^2 } - 2\:C \label{eq:cont_disp}
\end{equation}
we obtain the \emph{kinetic mass}
\begin{equation}
 M = M_{\rm kin} \equiv \frac{\bm{p}^2-\left[E(\bm{p})-E(0)\right]^2}{2\left[E(\bm{p})-E(0)\right]}. \label{eq:mkin}
\end{equation}
With the improved lattice actions used in this work, the continuum dispersion relation
(\ref{eq:cont_disp}) was found to be an excellent approximation for bottomonium at
small lattice momenta $\bm{p}$. This will be demonstrated in Sec.~\ref{sec:speed_of_light}.

\subsubsection{Bootstrap method}

When computing quantities that depend on more than one fit result, such as the kinetic mass (\ref{eq:mkin}) or an
energy splitting obtained from independent fits, correlations between
the different fit results must be taken into account. In this work, the bootstrap method
was employed to achieve this. Note however that it must be modified for Bayesian
fitting \cite{Lepage:2001ym} so that not only the data sets are resampled randomly but also
the central values $\tilde{p}_i$ of the priors in (\ref{eq:priors})
are drawn from Gaussian random distributions with widths $\sigma_{\tilde{p}_i}$ for every fit.

The final quantity of interest is then computed for the bootstrap ensemble of
fit results, giving an approximate probability distribution. In the end, the mean value and
the 68\% width of this distribution are quoted.

The bootstrap method was in fact not only used for quantities depending on
more than one fit parameter but also to obtain the error estimates for individual fit parameters.
The number of bootstrap samples was taken to be 500.

\subsubsection{\label{sec:autocorr}Autocorrelations}

The integrated autocorrelation time $\tau_{\rm int}$ for the 12th time slice of the pion correlator on
the $am_l=0.005$ ensemble was found to be 10 to 15 steps of molecular dynamics (MD) time in \cite{Allton:2008pn}.
Therefore only gauge configurations separated by 25 steps, which is approximately equal to $2\tau_{\rm int}$,
were used here. However, the autocorrelation time depends on the observable. The measurements in
this work were checked for residual autocorrelations using the binning method.

Recall from Sec.~\ref{sec:2ptcalc} that on each gauge configuration meson correlators from 32 origins
(8 origins on 4 source time slices each) were computed. The data was always averaged over the 8 origins
on each source time slice, thus leaving 4 data samples per configuration.

To estimate the autocorrelations between the different gauge configurations, the data was also averaged
over the 4 source time slices prior to the binning. Note that the measurements already had an initial
separation of 25 steps in MD time, and hence the binning increases this to integer multiples of 25.

Of course the binning reduces the number of data samples available for the fit. To obtain
a reliable estimate of the data covariance matrix (see e.g.~\cite{Lepage:2001ym}),
the number of data samples should be much larger than the dimension of this matrix. Thus,
in the analysis of autocorrelations, fits with only one smearing at source and sink
and a small fitting range, corresponding to a small data covariance matrix, were considered.

No significant increases in the bootstrap errors were seen for any of the ensembles, indicating 
that the separation of 25 steps of MD time gives sufficiently independent measurements.
Note that the origins of the meson correlators were shifted randomly between gauge
configurations.

Next, tests for autocorrelations between the data samples from the four different source time slices were performed.
To this end the data was averaged into bins of 2 or 4 time slices, without additional binning over
gauge configurations. Here, in some cases a slight increase in the bootstrap errors was seen, at most
20\%. Thus, for the measurements in the remainder of this work the following conventions were used:
on the $a m_l=0.005$ and $a m_l=0.01$ ensembles all 4 source time slices were binned together,
except for the $3\times 3$ matrix correlator in the determination of the $\Upsilon(3S)$ energy.
For the latter no binning over source time slices was done; instead the error estimates from the fits
were corrected by 20\% upwards to be safe.
For the $a m_l=0.02$ and $a m_l=0.03$ ensembles, which have about four times fewer configurations,
the $\Upsilon(3S)$ state was not computed. There, for the $2\times2$ matrix correlators
no binning over source time slices was performed, again increasing the error estimates by 20\%
upwards instead.
For the $D$-wave correlators, only the $1D$ smearing function was included in the fits. Thus,
the data covariance matrix was small and binning over all 4 source time slices was used.

\section{\label{sec:results}Results}

\subsection{\label{sec:mass_tuning}Tuning the bare $b$ quark mass}

The bare $b$ quark mass, which is a free parameter in the NRQCD action, was tuned non-perturbatively.
It was adjusted such that the kinetic mass of the $\eta_b(1S)$ meson as calculated on the lattice matches the experimental value
of $9.389(5)$ GeV \cite{Aubert:2008vj}. The tuning was done on the most chiral
($am_l=0.005$) ensemble of gauge configurations.

The kinetic mass was computed from (\ref{eq:mkin}), where the smallest possible
lattice momentum $a|\bm{p}|=1\cdot 2\pi /L$ was used. As shown in the next section,
the kinetic mass is very stable and shows no significant dependence on $\bm{p}$ even
for much larger momenta. In order to increase statistics the results were averaged
over the different possibilities for the direction of $\bm{p}$.

The comparison with experiment of course requires the knowledge of the lattice spacing, which
was determined as the ratio of the experimentally measured $\Upsilon(2S)-\Upsilon(1S)$ mass
splitting, $0.56296(40)$ GeV \cite{Amsler:2008zzb}, to the dimensionless lattice result.
This will be discussed in more detail in Sec.~\ref{sec:radial_orbital_splittings}.

The lattice results for $a M_{\rm kin}$ and the $\Upsilon(2S)-\Upsilon(1S)$ splitting at the three different
bare quark masses $a m_b=2.30,\:\:2.45$ and $2.60$ are shown in Table \ref{tab:mb_tuning}.
\begin{table}
\begin{tabular}{ccccccc}
\hline\hline
$am_b$ & \hspace{2ex} & $aM_\textrm{kin}(\eta_b)$ & \hspace{2ex} &  $\displaystyle\begin{array}{c}
\Upsilon(2S)-\Upsilon(1S) \\ \mathrm{splitting} \end{array}$ \\
\hline
$2.30$  &&  $4.988(12)$  &&  $0.3258(47)$ \\
$2.45$  &&  $5.281(13)$  &&  $0.3242(46)$ \\
$2.60$  &&  $5.575(13)$  &&  $0.3231(54)$ \\
\hline\hline
\end{tabular}
\caption{\label{tab:mb_tuning}Results for the tuning of the bare $b$ quark mass in lattice units. Errors are statistical/fitting only.}
\end{table}
As can be seen, the $\Upsilon(2S)-\Upsilon(1S)$ splitting is very insensitive to the value
of the $b$ quark mass. It is also expected to have much smaller lattice discretization errors
than the $1P-1S$ splitting as discussed in the next section.

\begin{figure}
\includegraphics[width=0.9\linewidth]{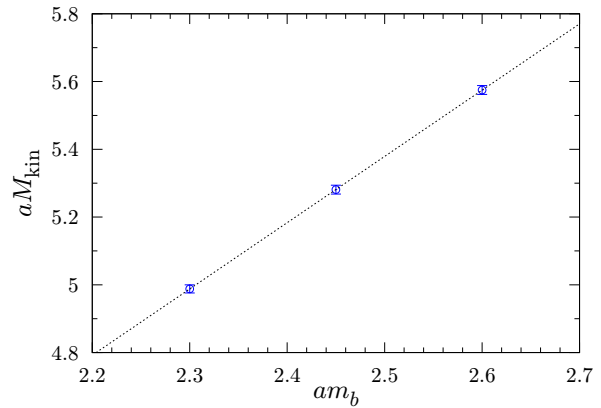}
\caption{\label{fig:mkin_vs_mb} The kinetic mass of the $\eta_b(1S)$ meson plotted against the bare heavy quark mass.
Errors are statistical/fitting only. The line shows the average over the bootstrap ensemble of linear fit results. }
\end{figure}

It was found that in the range considered here, the dependence of the kinetic mass on the bare heavy
quark mass is described very well by the linear relation
\begin{equation}
aM_\mathrm{kin}=A+B\cdot am_b. \label{eq:mkin_vs_mb}
\end{equation}
A plot of $aM_\textrm{kin}$ as a function of $am_b$ is
shown in Fig.~\ref{fig:mkin_vs_mb}.
Fits of Eq.~(\ref{eq:mkin_vs_mb}) with the parameters $A$ and $B$ were performed
on 500 bootstrap samples for the kinetic masses at $a m_b=2.30,\:\:2.45$ and $2.60$.
The resulting average fit parameters were
\begin{eqnarray}
\nonumber A & = & 0.489(25), \\
B & = & 1.956(11).
\end{eqnarray}
To obtain a first result for the lattice spacing of the $a m_l=0.005$ ensemble,
the $\Upsilon(2S)-\Upsilon(1S)$ mass splitting at $a m_b=2.45$ was used, giving $a^{-1}=1.736(25)$ GeV
(the error is statistical/fitting only). Of course the $b$ quark mass was not yet tuned,
but given the relative independence of the $\Upsilon(2S)-\Upsilon(1S)$ splitting on $m_b$, the
value of $a m_b=2.45$ was sufficiently close to the physical value. The final
results for the lattice spacing obtained with the correct $b$ quark mass will be presented
in Sec.~\ref{sec:radial_orbital_splittings}.

Using the preliminary result for $a^{-1}$, it follows that the $\eta_b(1S)$ mass in lattice units
must be tuned to be $a M_\mathrm{kin}=5.407(77)$. Inserting this into (\ref{eq:mkin_vs_mb})
and solving for $a m_b$ gives
\begin{equation}
a m_b = 2.514(36).  \label{eq:amb}
\end{equation}
The error quoted here is statistical/fitting only and is dominated by the
uncertainty in the lattice result for the $\Upsilon(2S)-\Upsilon(1S)$ splitting.

All remaining calculations were actually performed with $am_b=2.536$. This was an earlier result and
the fits have been improved slightly since then. However it is still inside the range of the new value (\ref{eq:amb}).

For $am_b=2.536$ the results were $aM_{\rm kin}=5.449(13)$ and $a^{-1}=1.740(25)$ GeV.
This gives $M_{\rm kin}=9.48(14)$ GeV which is compatible with the experimental result of
$9.389(5)$ GeV, confirming the successful tuning of the heavy quark mass.

Note that the lattice NRQCD action can be used for both heavy-heavy mesons and heavy-light hadrons.
Thus, the result for the bare $b$ quark mass obtained here will be useful also in future calculations
for heavy-light hadrons.

\subsection{\label{sec:speed_of_light}Speed of light}

In order to examine how well the lattice data approximates the relativistic continuum dispersion
relation (\ref{eq:cont_disp}), the kinetic mass of the $\eta_b(1S)$ meson, defined by (\ref{eq:mkin}),
was also computed for larger lattice momenta $a\bm{p}=\bm{n}\cdot 2\pi/L$ up to $\bm{n}^2=12$.
For these calculations, the local smearing function $\Gamma(\bm{r})=\delta_{\bm{r},0}$ was used
at source and sink so that multiple lattice momenta can be obtained with little computational cost.
For each value of $\bm{n}^2$, the results were averaged over the possible directions of the vector
$\bm{n}$, and all components of $\bm{n}$ were chosen to be less than or equal to $2$.

\begin{table}
\begin{tabular}{lllllll}
\hline\hline
$\bm{n}^2$ & \hspace{2ex} & $aM_\textrm{kin}(\eta_b)$ & \hspace{2ex} & $aC$          & \hspace{2ex} & $c^2$         \\
\hline
1          && $5.450(17)$       && $2.5913(84)$ && -             \\
2          && $5.450(17)$       && $2.5912(85)$ && $1.00003(85)$ \\
3          && $5.450(18)$       && $2.5911(92)$ && $1.0001(16)$ \\
4          && $5.461(22)$       && $2.597(11)$  && $0.9981(21)$  \\
5          && $5.457(20)$       && $2.595(10)$  && $0.9987(24)$  \\
6          && $5.452(20)$       && $2.592(10)$  && $0.9997(27)$  \\
8          && $5.454(22)$       && $2.593(11)$  && $0.9993(35)$  \\
9          && $5.447(20)$       && $2.590(10)$  && $1.0005(35)$  \\
12         && $5.445(21)$       && $2.589(11)$  && $1.0009(42)$  \\
\hline\hline
\end{tabular}
\caption{\label{tab:speed_of_light}Kinetic mass, NRQCD energy shift and the square
of the ``speed of light'' for various lattice momenta $a\bm{p}=\bm{n}\cdot 2\pi/L$,
calculated on the $a m_l=0.005$ ensemble with $a m_b = 2.536$.}
\end{table}

The results are given in Table \ref{tab:speed_of_light}, where also the NRQCD energy shift,
calculated as
\begin{equation}
C=\frac{M_{\rm kin}(\bm{p})-E(0)}{2},
\end{equation}
and for $\bm{n}^2>1$ the square of the ``speed of light''
\begin{equation}
 c^2 \equiv \frac{\left[E(\bm{p})-E(0)+M_{\rm kin,1}\right]^2-M_{\rm kin,1}^2}{\bm{p}^2} \label{eq:speed_of_light}
\end{equation}
are shown. In Eq.~(\ref{eq:speed_of_light}), $M_{\rm kin,1}$ denotes the kinetic mass calculated with $\bm{n}^2=1$.
In the units used here, one should have $c^2=1$. Deviations of $c^2$ from 1 can be caused by discretization errors
in the NRQCD, gluon and sea quark actions and also by missing higher order relativistic corrections in the
NRQCD action. The NRQCD action is highly improved at tree level, and so the most significant errors one expects
here are those caused by missing radiative corrections.

As can be seen in the table, in the momentum range considered here the kinetic mass shows no significant
dependence on $\bm{p}$ within the small statistical/fitting errors. Correspondingly, $c^2$ remains
compatible with 1, with statistical/fitting errors less than $0.5$\%, indicating that the effect of
the errors mentioned above is small.

Analogous calculations for the $\Upsilon(1S)$ meson have been performed in \cite{Gray:2005ur} with the
same NRQCD action but with the L\"{u}scher-Weisz gluon and the AsqTad sea quark action. There, the
deviation of $c^2$ from 1 in the same momentum range was also found to be compatible with 1 within statistical
errors of less than 1\%.

\subsection{\label{sec:radial_orbital_splittings}Radial/orbital splittings and the lattice spacing}

The lattice results for the various radial and orbital energy splittings are listed in Table
\ref{tab:radial_orbital_splittings}.

\begin{table*}
\begin{ruledtabular}
\begin{tabular}{lllll}
                                                        & $am_l=0.005$   & $am_l=0.01$  & $am_l=0.02$  & $am_l=0.03$  \\
\hline
$\Upsilon(2S)-\Upsilon(1S)$                             &  $0.3236(46)$  & $0.3270(73)$ & $0.330(18)$  & $0.327(23)$  \\
$\Upsilon(3S)-\Upsilon(1S)$                             &  $0.517(21)$   & $0.537(23)$  & -            & -            \\
\hline
$\langle\chi_{b}(1P)\rangle-\Upsilon(1S)$               &  $0.2589(30)$  & $0.2572(22)$ & $0.2628(57)$ & $0.2613(61)$ \\
$\langle\chi_{b}(2P)\rangle-\Upsilon(1S)$               &  $0.478(30)$   & $0.502(26)$  & $0.511(39)$  & $0.516(37)$  \\
$\langle\chi_{b}(2P)\rangle-\langle\chi_{b}(1P)\rangle$ &  $0.219(29)$   & $0.245(24)$  & $0.248(35)$  & $0.255(33)$  \\
\hline
$\Upsilon_2(1D)-\Upsilon(1S)$                           &  $0.4080(46)$  & $0.4194(42)$ & $0.417(12)$  & $0.426(12)$  \\
\end{tabular}
\end{ruledtabular}
\caption{\label{tab:radial_orbital_splittings}Results for the radial and orbital energy splittings in lattice units.
Errors are statistical/fitting only.}
\end{table*}

\begin{table*}
\begin{ruledtabular}
\begin{tabular}{lllll}
                                      & $am_l=0.005$      &  $am_l=0.01$    & $am_l=0.02$     & $am_l=0.03$     \\
\hline
$a^{-1}_{2S-1S}$ \hspace{1ex} (GeV)   &  $1.740(25)(19)$  & $1.722(38)(19)$ & $1.708(92)(19)$ & $1.72(12)(2)$   \\
$a^{-1}_{1P-1S}$ \hspace{1ex} (GeV)   &  $1.698(19)(65)$  & $1.709(15)(65)$ & $1.673(36)(64)$ & $1.682(40)(64)$ \\
\end{tabular}
\end{ruledtabular}
\caption{\label{tab:lattice_spacing}Results for the inverse lattice spacing obtained from the $\Upsilon(2S)-\Upsilon(1S)$
and $\langle\chi_{b}(1P)\rangle-\Upsilon(1S)$ splittings. The first error given is statistical/fitting and the
second is an estimate of the systematic errors (relativistic, radiative and discretization) due to the NRQCD action.
Systematic errors due to the gluon and sea quark actions are not included.}
\end{table*}

Systematic errors are known to be smallest for the spin-averaged masses, defined as
\begin{equation}
\langle M \rangle=\frac{\sum_J (2J+1)M_J}{\sum_J (2J+1)}. \label{eq:spinav}
\end{equation}
However, in most cases not all of the states entering Eq.~(\ref{eq:spinav}) are known from experiment.
For the $1S$, $2S$ and $3S$ masses in this section the $J=1$ states ($\Upsilon$) are considered
instead of the spin-averages. Note that the $J=0$ $S$-wave states ($\eta_b$) enter the spin-averaged
masses only with a weight of $1/4$, and so the influence of systematic errors
in the hyperfine splittings is negligible here.
For the $1P$ and $2P$ masses, the spin-averages over the $\chi_b$ triplet ($J=0,1,2$) states are used.
The only experimentally known $D$-wave state \cite{Bonvicini:2004yj} is $\Upsilon_2(1D)$ with $J=2$,
and this state is therefore considered here.

In terms of the NRQCD power counting \cite{Lepage:1992tx}, radial and orbital energy splittings are
of order $\mathcal{O}(v^2)$, where $v$ is the internal speed of the $b$ quarks inside the heavy-heavy meson.
For bottomonium one has $v^2\approx0.1$. The NRQCD action in use includes all relativistic corrections of order
$\mathcal{O}(v^4)$ (at tree-level), and hence the missing relativistic corrections are of order $\mathcal{O}(v^6)$. Naively
this leads to relativistic errors for the radial and orbital splittings of $\mathcal{O}(v^4)=1\%$. However, as
discussed in \cite{Gray:2005ur}, for energy splittings one has to consider the \emph{difference} between the
expectation values of the missing operators for the two states. This leads to a reduction of the relativistic errors
for the $2S-1S$ splitting to about $0.5$\%.

Additional systematic errors for the NRQCD action are due to discretization errors
and missing radiative corrections (beyond tadpole improvement). Estimates of these
errors for the $2S-1S$ and $1P-1S$ splittings are given in Table \ref{tab:NRQCD_syst_errors}.
\begin{table}
\begin{tabular}{lcccc}
\hline\hline
          & \hspace{2ex} &  $2S-1S$  & \hspace{2ex} & $1P-1S$ \\
\hline
  relativistic            &&  $0.5$\%  && $1.0$\%  \\
  radiative               &&  $0.5$\%  && $1.7$\%  \\
  discretization          &&  $0.8$\%  && $3.2$\%  \\
\hline
  total                   &&  $1.1$\%  && $3.8$\%  \\
\hline\hline
\end{tabular}
\caption{\label{tab:NRQCD_syst_errors}Estimates of the systematic errors due to the lattice NRQCD
action for the $2S-1S$ radial and $1P-1S$ orbital splittings \cite{Gray:2005ur}.}
\end{table}
They are taken to be equal to the estimates obtained in \cite{Gray:2005ur} for exactly the same
lattice NRQCD action on the ``coarse'' MILC gauge configurations, which have a lattice spacing
($a^{-1}\approx1.6$ GeV) very similar to the ensembles considered here. The reader
is referred to \cite{Gray:2005ur} and \cite{Davies:1998im} for the details.
As can be seen in the table, systematic errors are much smaller for the $2S-1S$ splitting
compared to the $1P-1S$ splitting. This is due to the smaller difference in the
wave functions for the $2S$ and $1S$ states. The $2S-1S$ splitting thus allows
a more reliable determination of the lattice spacing.

Note that there are also discretization errors due to the gluon and sea quark actions. These are difficult to
quantify at this stage as only data from one lattice spacing is available. Gauge configurations with a smaller
lattice spacing are currently being generated by the RBC and UKQCD collaborations so that a more systematic analysis
will become possible in the future.
In \cite{Allton:2008pn}, a preliminary error estimate of $(a\Lambda_{\rm QCD})^2\approx4$\% for the calculations of
light hadron properties on the current ensembles was given.
The calculations performed here are different in that the domain wall action only enters via the sea quarks.
The Iwasaki gluon action \cite{Iwasaki:1983ck,Iwasaki:1984cj} is renormalization-group-improved and is therefore
expected to have a better scaling behavior than the unimproved Wilson action. However, this depends on the
observable considered; see e.g.~\cite{Necco:2003vh} for a scaling study of the critical temperature and glueball masses.
The stability of the ``speed of light'' demonstrated in Sec.~\ref{sec:speed_of_light} provides some evidence for the
smallness of the effect of gluon discretization errors for bottomonium.

For reference, the discretization errors in the $2S-1S$ and $1P-1S$ splittings on the coarse MILC lattices
due to the L\"uscher-Weisz gluon action were estimated in \cite{Gray:2005ur} to be 0.5\% and 1.7\%, respectively.
These errors are proportional to the difference in the square of the wave function at the origin, which is
smaller between the $2S$ and $1S$ states.

The results for the inverse lattice spacings of the four ensembles from both the $\Upsilon(2S)-\Upsilon(1S)$
and the $\langle\chi_{b}(1P)\rangle-\Upsilon(1S)$ splittings are listed in Table \ref{tab:lattice_spacing}. For
the most chiral ensemble the $2S-1S$ splitting gives $a^{-1}=1.740(25)_{\rm stat}(19)_{\rm syst}$ GeV.
No significant dependence on the sea quark mass can be seen within the statistical errors,
and therefore no extrapolation was attempted.
For comparison, the RBC and UKQCD collaborations have obtained $a^{-1}=1.729(28)_{\rm stat}$ in the chiral limit, using
the $\Omega^{-}$ baryon mass \cite{Allton:2008pn}. This is consistent with the results obtained here.

\begin{figure}
\includegraphics[width=0.95\linewidth]{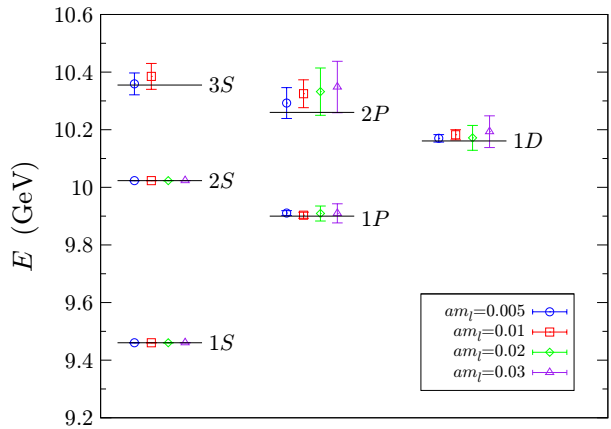}
\caption{\label{fig:radial_orbital_vs_experiment} Radial and orbital energy splittings compared to the
experimental results (indicated by lines). Errors are statistical/fitting only and include the uncertainty
in the determination of the lattice spacing. The $1S$ and $2S$ masses, for which no error bars are shown,
are not predictions of the lattice calculation as these states are used to determine the lattice scale
and the overall energy shift.}
\end{figure}

Next, the lattice spacing determinations from the $2S-1S$ splitting were used to convert the other radial
and orbital splittings from Table \ref{tab:radial_orbital_splittings} to physical units. The results are plotted
in Fig.~\ref{fig:radial_orbital_vs_experiment}.
Note that the individual results for the lattice spacings of the different ensembles were used.

Overall, good agreement with experiment is seen. The dependence on the light sea quark mass
is found to be weak. This is expected since the typical gluon momenta inside the bottomonium are
much larger than all the values for the light quark masses used here.
However, note that large deviations between lattice results and experiment were previously
seen in quenched simulations ($n_f=0$), so the inclusion of 2+1 flavors of dynamical light
quarks is in fact very important. A comparison between quenched and unquenched results
can be found in \cite{Davies:2003ik}.

\subsection{\label{sec:spin_dep_splittings}Spin-dependent energy splittings}

The spin-dependent energy splittings in bottomonium, i.e. the fine and hyperfine structure, are of order $\mathcal{O}(v^4)$
and hence any sub-leading corrections are missing in the NRQCD action used here. Therefore, the relativistic errors
in these splittings are expected to be of order $\mathcal{O}(v^2)\approx 10$\%. The spin-dependent energy splittings
also receive radiative corrections of order $\mathcal{O}(\alpha_s)$, the strong coupling constant at the scale set by
the lattice spacing. This leads to further systematic errors of the order of 20\%, although tadpole improvement
reduces the problem. Finally, discretization errors are also expected to be larger than for
the radial and orbital splittings, especially for the $S$-wave hyperfine splitting as discussed below.

The results for the spin-dependent energy splittings in lattice units are summarized in Table \ref{tab:spin_splittings}, where the errors
given are statistical/fitting only.

\begin{table*}
\begin{ruledtabular}
\begin{tabular}{lllll}
                                            & $\ms am_l=0.005$   & $\ms am_l=0.01$   & $\ms am_l=0.02$   & $\ms am_l=0.03$   \\
\hline
$\Upsilon(1S)-\eta_b(1S)$                   &  $\ms 0.03017(14)$ & $\ms 0.03033(16)$ & $\ms 0.03102(36)$ & $\ms 0.03145(38)$ \\
$\Upsilon(2S)-\eta_b(2S)$                   &  $\ms 0.0137(30)$  & $\ms 0.0120(48)$  & $\ms 0.013(12)$   & $\ms 0.018(16)$   \\
\hline
$\chi_{b0}(1P)-\langle\chi_{b}(1P)\rangle$  &     $-0.0207(20)$  &    $-0.0206(18)$  &    $-0.0231(36)$  &    $-0.0175(70)$  \\
$\chi_{b1}(1P)-\langle\chi_{b}(1P)\rangle$  &     $-0.0049(14)$  &    $-0.0027(19)$  &    $-0.0059(22)$  &    $-0.0049(41)$  \\
$\chi_{b2}(1P)-\langle\chi_{b}(1P)\rangle$  &  $\ms 0.0071(11)$  & $\ms 0.0058(12)$  & $\ms 0.0082(17)$  & $\ms 0.0064(29)$  \\
$h_b(1P)-\langle\chi_{b}(1P)\rangle$        &     $-0.0026(18)$  &    $-0.0002(21)$  &    $-0.0014(27)$  &    $-0.0058(42)$  \\
$\chi_{b1}(1P)-\chi_{b0}(1P)$               &  $\ms 0.0158(18)$  & $\ms 0.0176(25)$  & $\ms 0.0173(40)$  & $\ms 0.0126(77)$  \\
$\chi_{b2}(1P)-\chi_{b1}(1P)$               &  $\ms 0.0120(23)$  & $\ms 0.0088(31)$  & $\ms 0.0137(38)$  & $\ms 0.0113(68)$  \\
$h_b(1P)-\chi_{b1}(1P)$                     &  $\ms 0.0023(16)$  & $\ms 0.0027(16)$  & $\ms 0.0044(35)$  &    $-0.0009(61)$  \\
\hline
$\Upsilon_2(1D)-\eta_b(1D)$                 &  $\ms 0.0011(21)$  &    $-0.0012(18)$  &    $-0.0086(70)$  &    $-0.0050(61)$ \\
\end{tabular}
\end{ruledtabular}
\caption{\label{tab:spin_splittings}Spin-dependent energy splittings in lattice units. Errors are
statistical/fitting only. Large systematic errors are expected as discussed in the text.}
\end{table*}

\subsubsection{$S$-wave hyperfine structure}

\begin{figure}
\includegraphics[width=0.95\linewidth]{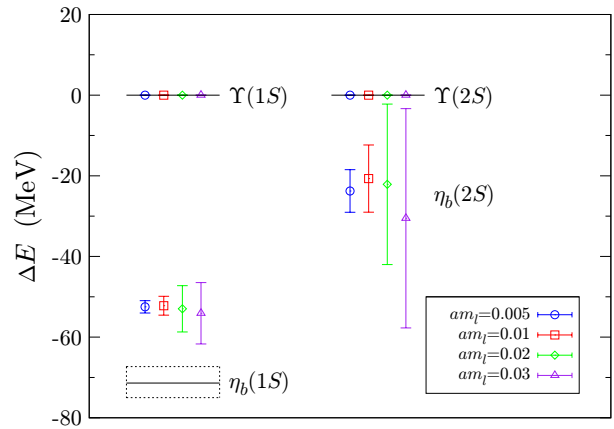}
\caption{\label{fig:S_wave_hyperfine} S-wave hyperfine splittings (energies relative to the
$\Upsilon(1S)$ and $\Upsilon(2S)$ states, respectively) compared to experiment. Errors are statistical/fitting
only and include the uncertainty in the determination of the lattice spacing, which enters with a factor of 2.
Large systematic errors are expected as discussed in the text.}
\end{figure}

Figure \ref{fig:S_wave_hyperfine} shows a plot of the $\Upsilon(1S)-\eta_b(1S)$ and
$\Upsilon(2S)-\eta_b(2S)$ energy splittings, where the previous lattice spacing
determinations from the $2S-1S$ splittings were used to convert to physical units.

The errors shown are statistical/fitting only but include the uncertainty
in the determination of the lattice spacing. The latter in fact enters with a factor of 2 here, as discussed in
\cite{Davies:1998im}, due to the resulting uncertainty in the physical heavy quark mass
(the hyperfine splitting is approximately proportional to the inverse of that mass).
The statistical error in the $1S$ hyperfine splitting is then dominated by far by this uncertainty, while
the $2S$ hyperfine splitting has an intrinsically higher statistical error as the state is radially excited.

The $\Upsilon(1S)-\eta_b(1S)$ splitting has recently been measured by the BABAR collaboration
\cite{Aubert:2008vj}, who found $71.4^{+2.3}_{-3.1}({\rm stat})\pm2.7({\rm syst})$ MeV. This value
is indicated in Fig.~\ref{fig:S_wave_hyperfine}. The lattice result in physical units for the
$am_l=0.005$ ensemble is $52.5\pm1.5({\rm stat})$ MeV, which is too small by about 25\%,
in line with the large systematic errors expected. Similarly to the radial and orbital splittings,
little dependence on the light sea quark mass is seen, which is expected for the same reason as discussed there.

Note that in \cite{Gray:2005ur} and \cite{Davies:1998im} a significant dependence
on the lattice spacing was found, with the result increasing toward
finer lattices, indicating that a substantial part of the deviation is due to discretization
errors. The hyperfine splitting is indeed expected to be sensitive to very short distances,
as the spin-spin interaction potentials in simple models contain a delta function at
the origin (see e.g. \cite{Buchmuller:1992zf}).

Finally, note that in \cite{Li:2008kb}, where a relativistic heavy-quark action was used, the
$\Upsilon(1S)-\eta_b(1S)$ splitting on the same RBC/UKQCD gauge configurations
was found to be only $23.7\pm3.7({\rm stat})$ MeV, a much larger deviation to experiment
than found here.

\subsubsection{$P$-wave spin-dependent splittings}

\begin{figure}
\includegraphics[width=0.95\linewidth]{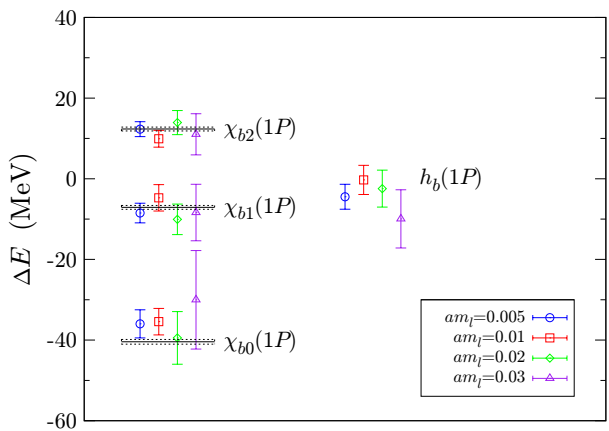}
\caption{\label{fig:P_wave_spin_splittings} P-wave spin splittings (energies relative to the
spin-average of the $\chi_{b}(1P)$ states) compared to experiment. Errors are statistical/fitting only and
include the uncertainty in the determination of the lattice spacing. Large systematic errors are expected
as discussed in the text.}
\end{figure}

A plot of the $1P$ spin-dependent splittings, converted to physical units using the previous
$2S-1S$ lattice spacing results, is given in Fig.~\ref{fig:P_wave_spin_splittings}.

It shows the energy differences of the $\chi_{b0}(1P)$, $\chi_{b1}(1P)$, $\chi_{b2}(1P)$ and $h_b(1P)$
states to the spin-average of the triplet $\langle \chi_b(1P) \rangle$. The experimental results 
\cite{Amsler:2008zzb} for the triplet states are also indicated in the plot; the $h_b$ states have not
yet been observed.

The lattice results are found to be in relatively good agreement with experiment, even within the purely statistical/fitting
errors shown in the plot (those include the uncertainty in the lattice spacing). This indicates that discretization
errors may be smaller than for the $S$-wave hyperfine splittings. Note that in simple potential models the
wave function at the origin is zero (cf. the smearing functions in Table \ref{tab:smear_funcs})
and hence the $P$-wave spin splittings are expected to be not as sensitive to very short distances
as the $S$-wave hyperfine splittings.

The result for the experimentally unknown $h_b(1P)-\langle\chi_{b}(1P)\rangle$ splitting on the $am_l=0.005$ ensemble
is $-4.5\pm3.1$ MeV, where the error quoted is statistical/fitting only and includes the uncertainty
from the determination of the lattice spacing.

\subsubsection{$D$-wave spin-dependent splittings}

Here, only the $\Upsilon_2(1D)-\eta_b(1D)$ splitting was calculated using the $E^{--}$ and $T_2^{-+}$ representations,
as these two states do not mix and can be computed from the same heavy-quark propagators.

The lattice results for the different ensembles are listed in Table \ref{tab:spin_splittings}.
On the $am_l=0.005$ ensemble, the splitting in physical units is found to be $1.8\pm3.7$ MeV
where the error given is statistical/fitting only and includes the uncertainty from the
determination of the lattice spacing. No experimental results are available.

\section{\label{sec:discussion}Conclusion}

In this paper, a comprehensive calculation of the bottomonium spectrum with improved lattice NRQCD
on the RBC/UKQCD $24^3\times 64$, $L_s=16$ gauge configurations with 2+1 flavors of dynamical
domain wall fermions was presented. The results are similar to those obtained in \cite{Gray:2005ur}
with the same heavy-quark action from the ``coarse'' MILC gauge configurations, which use the AsqTad
fermion action and the L\"{u}scher-Weisz gluon action. In particular, good agreement with experiment
was found for the radial and orbital energy splittings, for which systematic errors due to the NRQCD action
are small. Furthermore, no significant deviations of the ``speed of light'' from 1 in the $\eta_b(1S)$
dispersion relation were found within the small statistical errors. The calculations in this work
provide further evidence for the good properties of the domain wall and Iwasaki actions employed
by the RBC and UKQCD collaborations. By comparing the $2S-1S$ radial energy splitting to experiment,
independent determinations of the lattice spacings were performed, giving
$a^{-1}=1.740(25)_{\rm stat}(19)_{\rm syst}$ GeV for the most chiral ensemble.

The results for the fine and hyperfine structure are expected to have larger systematic errors due to
missing radiative and relativistic corrections in the NRQCD action as well as discretization errors.
Nevertheless, relatively good agreement with experiment was seen for the $P$-wave fine structure,
and the deviation to experiment in the $1S$ hyperfine splitting was found to be much smaller than
for the previous result obtained from a relativistic heavy quark action in \cite{Li:2008kb}.

The calculations presented here are only for one lattice spacing. A more systematic analysis of
discretization errors will be performed once new ensembles with a finer lattice spacing are made
available by the RBC and UKQCD collaborations.

Having obtained the bottomonium spectrum and the bare heavy quark mass in this work, the next step
is to perform calculations for heavy-light systems. Results for the spectrum of heavy-light baryons
and mesons with domain wall valence quarks but with \emph{static} heavy quarks were recently
presented in \cite{Detmold:2008ww}. Heavy-light computations with NRQCD heavy- and domain wall
light valence quarks on the same RBC/UKQCD gauge field ensembles are currently underway.

\begin{acknowledgments}

The author would like to thank Matthew Wingate for initially suggesting this
project, for many valuable discussions and for reading the manuscript.
Thanks also go to Christine Davies and Ron Horgan for helpful discussions.
The author is indebted to the RBC and UKQCD collaborations for making their gauge configurations publicly available.
The gauge configurations were fixed to Coulomb/Landau gauge using the Chroma software \cite{Edwards:2004sx}.
The numerical calculations in this work were performed using the Darwin Supercomputer of the University of Cambridge
High Performance Computing Service (\url{http://www.hpc.cam.ac.uk/}), provided by Dell Inc.~using
Strategic Research Infrastructure Funding from the Higher Education Funding Council for England.
Access to the Darwin Supercomputer was provided by Ron Horgan.
The author was supported by St John's College Cambridge, The Cambridge European Trust and EPSRC.

\end{acknowledgments}

\end{document}